\documentclass{tcibook}
\usepackage{fancyhea}
\usepackage{sidecap}
\usepackage{wrapfig}
\usepackage{framed}
\usepackage{work}
\usepackage{bm}       
\usepackage{graphicx}
\usepackage{hyperref}      
\usepackage{chngcntr}
\usepackage{hyperref}
\usepackage{lineno}

\counterwithout{figure}{chapter}

\newcommand{\nc}{\newcommand}  

\def\beq{\begin{equation}}
\def\eeq#1{\label{#1}\end{equation}}
\def\eeqn{\end{equation}}

\newenvironment{Eqnarray}%
   {\arraycolsep 0.14em\begin{eqnarray}}{\end{eqnarray}}
\def\beqa{\begin{Eqnarray}}
\def\eeqa#1{\label{#1}\end{Eqnarray}}
\def\eeqan{\end{Eqnarray}}

\nc{\ra}{\rightarrow}  
\nc{\slsh}{\slash\hspace*{-0.22cm}}
\def\Re{{\cal R \mskip-4mu \lower.1ex \hbox{\it e}\,}}
\def\Im{{\cal I \mskip-5mu \lower.1ex \hbox{\it m}\,}}

\nc{\vev}[1]{ \left\langle {#1} \right\rangle }
\nc{\bra}[1]{ \langle {#1} | }
\nc{\ket}[1]{ | {#1} \rangle }
\nc{\fb}{\,{\rm fb}^{-1}}
\nc{\ev}{{\rm eV}}
\nc{\kev}{{\rm keV}}
\nc{\Mev}{{\rm MeV}}
\nc{\gev}{{\rm GeV}}
\nc{\tev}{{\rm TeV}}
\nc{\mev}{{\rm MeV}}

\def\del{\partial}
\def\Dslash{\not{\hbox{\kern-4pt $D$}}}
\def\dslash{\not{\hbox{\kern-2pt $\del$}}}
\def\pslash{\not{\hbox{\kern-2pt $p$}}}
\def\ETmiss{ \not{\hbox{\kern-4pt $E$}}_T }

\def\msb{{\bar{\ssstyle M \kern -1pt S}}}

\begin{document}

\def\bibname{References}

\bibliographystyle{utphys}  

\raggedbottom

\pagenumbering{roman}

\parindent=0pt
\parskip=8pt
\setlength{\evensidemargin}{0pt}
\setlength{\oddsidemargin}{0pt}
\setlength{\marginparsep}{0.0in}
\setlength{\marginparwidth}{0.0in}
\marginparpush=0pt

\pagenumbering{arabic}
\renewcommand{\arraystretch}{1.25}
\addtolength{\arraycolsep}{-3pt}


\newcommand{\sfig}[2]{
\centering
\includegraphics[width=#2]{#1}
}
\newcommand{\Sfig}[2]{
    \begin{figure}[htbp]
    \sfig{#1.pdf}{0.6\columnwidth}
   \caption{#2} 
    \label{fig:#1}
    \end{figure}
}
\newcommand{\Sfigtwo}[3]{
    \begin{figure}[htbp]
    \sfig{#1.pdf}{0.45\columnwidth}
    \sfig{#2.pdf}{0.45\columnwidth}
   \caption{#3} 
    \label{fig:#1}
    \end{figure}
}

\newcommand{\rf}[1]{~\ref{fig:#1}}
 
\chapter*{Cosmic Visions Dark Energy: Science}

\begin{center}\begin{boldmath}

Scott Dodelson, Katrin Heitmann, Chris Hirata, Klaus Honscheid, Aaron Roodman, Uro\v{s} Seljak, An\v{z}e Slosar, Mark Trodden


\end{boldmath}\end{center}

\section*{Executive Summary}

Cosmic surveys provide crucial information about high energy physics including strong evidence for dark energy, dark matter, and inflation. Ongoing and upcoming surveys will start to identify the underlying physics of these new phenomena, including tight constraints on the equation of state of dark energy, the viability of modified gravity, the existence of extra light species, the masses of the neutrinos, and the potential of the field that drove inflation. Even after the Stage IV experiments, DESI and LSST, complete their surveys, there will still be much information left in the sky. 
This additional information will enable us to understand the physics underlying the dark universe at an even deeper level and, in case Stage IV surveys find hints for physics beyond the current Standard Model of Cosmology, to revolutionize our current view of the universe.
There are many ideas for how best to supplement and aid DESI and LSST in order to access some of this remaining information and how surveys beyond Stage IV can fully exploit this regime. These ideas flow to potential projects that could start construction in the 2020's.


\newpage
\renewcommand*\thesection{\arabic{section}}

\section{Overview}

This document begins with a description of the scientific goals of the cosmic surveys program in \S2 and then \S3 presents the evidence that, even after the surveys currently planned for the 2020's, much of the relevant information in the sky will remain to be mined. \S\ref{sec:extracting_info} presents a selection of ideas of how to extract this information. All of these flow into \S\ref{sec:project}, which lists several potential projects, each of which could implement at least some of the ideas in the previous section.

{\it This paper is intended to be informational for the Department of Energy (DOE) and does not imply any agency concurrence.}

\section{Physics} 


The universe began in a hot, dense state (the Big Bang), and expanded and cooled, initially going through a phase in which hot radiation was the dominant form of energy. 
As the expansion continued, the energies of radiation and matter became comparable, and the hot radiation then streamed unhindered through the cosmos, to be detected 13.8 billion years later by our telescopes as the Cosmic Microwave Background (CMB). As matter became dominant, the inexorable force of gravity grew minor imperfections in its distribution into larger and larger aggregations, ultimately forming the complex large-scale structure in the universe today.

This broad picture, based on Einstein's theory of gravity, General Relativity (GR), works only after addressing some difficult questions. What produced the initial imperfections or {\it inhomogeneities}? Why is the universe's expansion rate accelerating today? How did the inhomogeneities grow from the level of $10^{-4}$ imprinted in the CMB to become nonlinear today? A universe filled with ordinary matter and governed by GR would not produce these observations: the expansion would be decelerating and the inhomogeneities would still be linear. Previous generations of cosmic surveys and theoretical work have produced a model, invoking dark matter, dark energy, and inflation,  that addresses these questions and provides a remarkably successful account of the behavior of the universe from the earliest moments after the Big Bang all the way to the present. 

Like all good scientific stories, this one raises an entirely new set of questions: What is the underlying physics? What is the dark energy? What is the ultra-high energy theory responsible for inflation? What is the dark matter? 
%
The most powerful way to address these questions is to carry out cosmic surveys, in which large numbers of cosmological objects are measured, providing an array of data about the growth and evolution of structure in the universe.

The current epoch of cosmic acceleration seems to require one of three possible extensions to the Standard Model. One is that the cosmological constant (CC) is driving acceleration. This is a solution that would require the addition of a single new parameter to the Standard Model, but that raises longstanding questions for fundamental theory. There currently exists no clear theoretical explanation of a CC of the required magnitude. Another possibility is that the ultimate value of the cosmological constant is zero, and that cosmic acceleration is driven by the potential energy of a new field, with some mechanism to dynamically relax it to a small value. This notion naturally leads to forms of {\it dark energy} that invoke a slowly-rolling cosmological scalar field to source accelerated expansion. In such models it is difficult to see how to avoid an additional fine tuning at the same level as just tuning the bare CC. Nevertheless, these dynamical dark energy models offer the simplest extension of the CC model and some of them  
address the {\it coincidence problem} - why is the energy density in the accelerating component comparable with that in matter at this cosmic epoch?


Another, more radical, proposal is to maintain the standard set of particles and fields, but require that GR break down at very low energies/low densities/large scales, implying that it must be replaced with a new theory of {\it modified gravity}. 
Constructing viable modified gravity models has proven to be very challenging, as fatal instabilities and inconsistencies are often hidden in the simplest formulations.
Common to most models is the realization that any new gravitationally coupled degrees of freedom run the risk of being ruled out by local tests of gravity unless some new physics comes into play. Thus, we require mechanisms through which these new degrees of freedom are {\it screened} in the Solar System, and in other settings where most precision tests hold.

In the case of dark energy models, the equation of state parameter, $w$, and its derivative, $w'$ 
determine both the expansion history of the universe (the redshift-distance relation) 
and the rate at which inhomogeneities grow with time and scale. Modified gravity models decouple these two sets of predictions, so that {a modified gravity model with an expansion history identical to that predicted by a CC generically predicts a different rate for the growth of structure.} 
It is this difference that -- once the expansion history has been determined very precisely -- will distinguish the dark energy idea from modified gravity. 
The emphasis therefore will be not only be on precision measurements of $w$ and $w'$, but also of the more physical measurements of expansion history and growth of structure. 
\begin{framed}
{\bf The growth of structure measured at a number of different redshifts and length scales will distinguish between the two possible explanations for cosmic acceleration: GR+dark energy or modified gravity}.
\end{framed}

While surveys are the most appropriate tool with which to address the central question of cosmic acceleration, they are also extremely well-suited to probing physics that was dominant in the universe in a number of much earlier cosmic epochs, providing discovery potential for a range of new phenomena.

As emphasized during the Snowmass process, the most compelling explanation for the initial perturbations -- inflation -- requires new ultra-high energy physics and so offers a window onto scales that cannot be probed by accelerators. Current and upcoming survey measurements will allow tests of a range of early universe effects relevant to inflation, captured phenomenologically by a number of different parameters. Of particular interest for pinning down the physics are measurements of the scalar spectral index $n$, the running with scale of this index $\alpha\equiv dn/(d\log k)$, the amplitude of primordial nongaussianity, $f_{\rm NL}$, and the spatial curvature $\Omega_k$. The spectral index distinguishes among models. Generally these models predict that the running $\alpha$ is of order $(n-1)^2$, which has been measured to be $10^{-3}$, so measuring this small running 
would provide significant constraints on, or perhaps evidence against, inflation. 
The extent to which the universe is spatially flat ($\Omega_k=0$) is a central test of the inflationary paradigm itself, whereas measures of $f_{\rm NL}$ allow for the possibility of both falsifiability and model differentiation, since simple models do not predict significant nongaussianity. 

\begin{framed}
{\bf Surveys that probe dark energy will also provide valuable information about other pieces of fundamental physics.}
\end{framed}

Efforts at Snowmass brought to the fore another piece of physics probed by cosmic surveys. Neutrinos leave an imprint on the evolution of structure in the universe in a small but measurable way. These measurements are directly tied to quantities, such as the neutrino masses and the existence of sterile neutrinos, that can also be directly probed in terrestrial experiments. The same cosmic surveys that will help us solve the mysteries of the dark sector and inflation will also provide complementary information about neutrinos, information that will either support the 3-neutrino paradigm or upset it in ways that could reverberate through many sub-fields of physics. Key milestones over the next decade include: measurement of the sum of the neutrino masses, $\sum m_\nu$, and percent-level constraints on the effective number of relativistic species permeating the universe, $N_{\rm eff}$. While Snowmass emphasized that these latter constraints weigh in on the key question of the existence of light sterile neutrinos, recently it has become clear that they also constrain many models with {\it hidden sectors}. This is another area where cosmic surveys and the CMB complement one another, since information from both will be needed to obtain the percent level constraints that will be most valuable.

{\bf Dark Matter in Surveys:}  Astronomical observations have provided the only evidence to date for the existence of dark matter. Most quantitatively, they are sensitive to the amount of cosmological dark matter so constrain particle physics models such as supersymmetry that introduce particles that were thermally produced in the early universe. Upcoming surveys also may provide clues about the nature of the dark matter, but this is outside the scope of this report. We note here though that surveys currently provide the tightest constraints on warm dark matter and these are likely to improve in the future. Dwarf galaxies provide targets for indirect detection; these can be found in surveys and then followed up with more powerful telescopes to infer their mass profile and therefore tighten the predictions for indirect detection. There are other ideas for constraining the nature of dark matter and its interactions using the kinds of surveys considered here plus follow-up observations with other instruments.




\section{Information in the Sky}\label{information}

In particle physics, the new physics discovery potential of experiments depends, broadly speaking, on two factors; increasing the energy of collisions and increasing the luminosity of accelerators. As the LHC increases its energy, it will access new states not accessible at lower energy. As it increases its luminosity, the LHC will produce, for example, many more Higgs particles, and their various decay modes can be measured more and more accurately. Furthermore, increased luminosity means that even very rare decays 
may be discovered at the statistically significant level. If the partial widths to any of these channels 
disagree with those predicted by the Standard Model, we will have evidence for new physics. The key though is to generate enough interactions, or to produce enough Higgs particles, to enable these careful statistical determinations.

An analogous pair of considerations are relevant to the problem of extracting new physics from cosmic surveys. In this case, the analog of the unexplored frontier of higher energy is higher redshift and the analog of luminosity is the number of objects tracing the underlying density field. The Dark Energy Spectroscopic Instrument (DESI) and the Large Synoptic Survey Telescope (LSST) carry so much promise because they will both explore the universe out to higher redshift and detect many more objects than surveys that preceded them such as the Dark Energy Survey (DES) and the Sloan Digital Sky Survey (SDSS). These increases in both the ``energy'' and ``luminosity'' of the surveys provide complementary ways to increase their discovery potential. The analog of the small hints for new physics encoded in the decay rates is the shape of the power spectrum, which is a way to quantify inhomogeneities. Different types of dark energy or modified gravity leave different signatures in the shape of the power spectrum as a function of length scale or time. The same is true for the sum of the neutrino masses, the number of cosmic relativistic species, and various parameters that govern the initial perturbations from inflation.

By detecting even more objects or going to higher redshift than DESI or LSST, we can obtain more precise measures of the inhomogeneities, which in turn are sensitive to new physics. Figure~\rf{dpower} shows these projected improvements as a function of the number of objects detected for several different values of redshift. While DESI will obtain many objects at $z=1$, it will measure fewer at higher redshifts, so there will remain an enormous amount of information to be mined at $z>1$. To extract all of this information and learn as much as possible about fundamental physics, future surveys will have to measure more objects at redshift $z>1$.

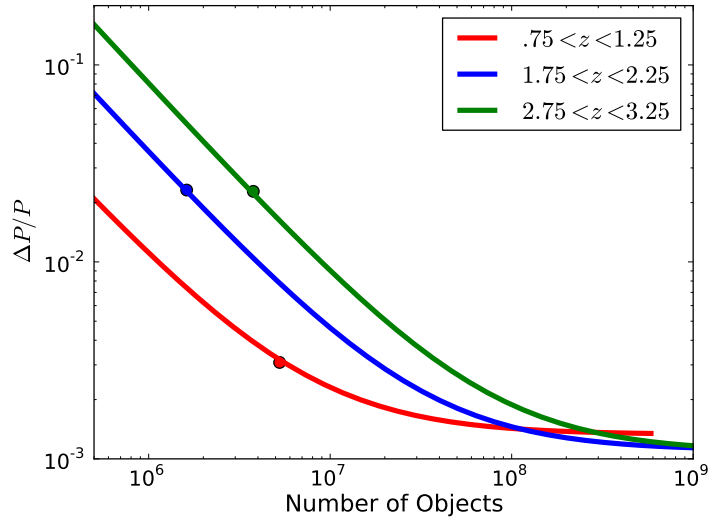
\Sfig{dpower}{Fractional error in the power spectrum on linear scales ($k=0.2h Mpc^{-1}$) that quantifies inhomogeneities for various redshifts as a function of the number of objects surveyed. The dots are projections for DESI: at $z=1$ DESI will be within a factor of 3 of the ultimate error, but at higher redshift, there is at least of factor of ten more information to be mined by future surveys. LSST will measure many more objects but will have imperfect radial information so therefore less effective information per object.}

There is a subtlety hidden in this analogy, one that argues even more strongly to go to higher redshift. The length scales on which structure is measured in the universe fall broadly into three categories. On large scales, those exceeding roughly 100 Mpc, the fluctuations are in the \emph{linear regime}, where our theoretical modeling is practically perfect and we can extract all information available in those modes. As surveys gather more objects over the same volume of the universe, we also begin to measure the \emph{weakly non-linear} scales between roughly 10 and 100 Mpc. On those scales, the modeling becomes more difficult, but not intractable and the signature of fundamental physics is not completely erased. On these smaller scales, there are many more weakly non-linear modes accessible for a survey (in Fourier space, the number of modes grows as $k^3$) and hence being able to extract information from them offers potentially massive rewards in terms of information content. This regime is an area of very active research in theoretical modeling and despite large advances in the field, we still do not know just how far into this regime we can model. Therefore it is impossible to make accurate forecasts for the scientific reach of the future experiments. Finally, on the smallest scales, those at separations less than approximately 10 Mpc, we find \emph{virialized structures}, where memory of the concrete realization of the initial conditions has been erased and remain there only in a statistical sense. These structures are the most difficult to model a-priori, but can be understood with a combination of numerical work and analytical models. Since in many models of modified gravity, the field goes from being screened to unscreened on these scales (see, e.g., \S\ref{novel}), they potentially offer novel ways of constraining these models. 

The future projects that probe higher redshifts thus extend our ability to constrain the universe in two distinct ways. First, by probing a different redshift range, they probe a different epoch in the evolution of the universe (as the LHC probes higher energy). For example, if dark energy had a non-standard behavior at $z\simeq1.5$, this can be established convincingly only by measuring the properties of the universe in the relevant redshift range. Second and equally important, reaching to higher redshift opens more volume, where a ``fresh'' batch of linear modes can be used for analysis. Moreover, since the universe was less evolved at those early times, there are more modes that had not yet gone nonlinear. 
Therefore, in terms of theoretical modeling, extending the redshift range is a conservative way of making sure we measure as much as we can using theoretical modeling we understand best.  As seen in Fig.~\rf{dpower}, there is only a factor of three to be gained from the linear regime by obtaining more objects than DESI at $z=1$. However, there will remain an enormous amount of information lurking at higher redshifts. 

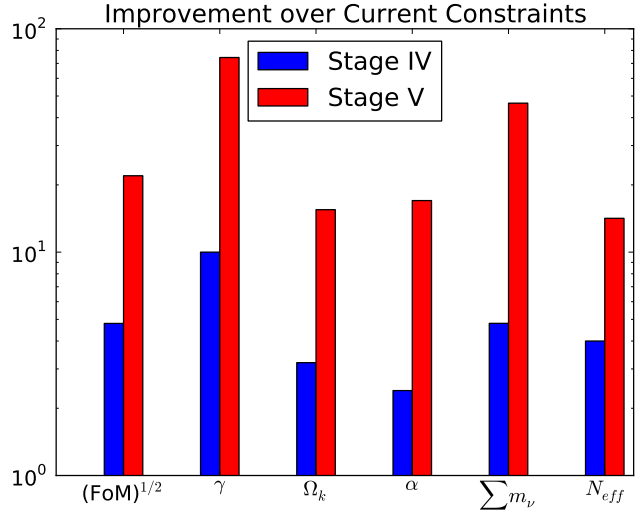
\Sfig{improvement}{Projected improvement (note the log scale) in constraints on parameters of dark energy (the Figure of Merit is inversely proportional to the allowed region in the $w,w'$ plane and $\gamma$ parametrizes the rate at which structure grows); inflation (curvature $\Omega_k$ and running $\alpha$); and neutrinos. The blue bar shows projected improvement over current constraints expected from the Stage IV experiments DESI and LSST, and the red bar shows improvements over current from a Stage V survey, indicating that there will still be large potential gains left even after the Stage IV surveys, DESI and LSST. {\it Details: Current constraints vary depending on which datasets are used. The current constraints used here come from a projection of Planck and BOSS data and are roughly equivalent to those in the Planck cosmological parameter paper. Projections for Stage IV assume Planck + DESI + LSST going out to scales $k=0.2$ h Mpc$^{-1}$. The projections for Stage V assume spectra for LSST galaxies and include information out to $k=0.5$ h Mpc$^{-1}$.}}

Figure~\rf{improvement} shows the potential of new surveys to discover new physics.
  We plot improvements in the parameter constraints going from current constraints to Stage IV dark energy experiments and then beyond to a Stage V experiment. This hypothetical experiment assumes big, but not irrational, improvements in modeling the large scale structure in the weakly nonlinear regime, but the projections should be taken not as accurate forecasts for any of the more concrete proposals that we advocate later in this report, but simply as a reflection on the amount of information that remains to be mined 
in the next few decades.

\begin{framed}
{\bf Even after the currently planned surveys finish operating, we can make revolutionary discoveries with future surveys; one indication of the power of these surveys is the projected order of magnitude improvements in parameter space.}
\end{framed}

\section{Extracting the Information}\label{sec:extracting_info}

Section 3 illustrated that, even after DESI and LSST, there will remain an enormous amount of information left in the sky.
Here we mention a representative subset of the many ideas for how to enhance DESI and LSST and ultimately move to the next stage of surveys.
This section is organized around the four probes first identified by the Dark Energy Task Force: Type Ia Supernovae (SN), Galaxy Clusters, Weak Gravitational Lensing, and Baryon Acoustic Oscillations (BAO) supplemented by Redshift Space Distortions (RSD). The first subsection highlights overarching ideas that benefit all of these probes; then, after a sub-section on promising future ideas for each probe, a concluding subsection revisits a topic that gained traction at Snowmass, Novel Probes of modified gravity.

\begin{framed}
{\bf There are many ideas for how to supplement DESI and LSST and  how to build future surveys that would access the remaining information in the sky.}
\end{framed}

Many of these ideas involve optical surveys. Optical surveys can broadly be categorized into two approaches: spectroscopic surveys, such as DESI, and photometric imaging surveys, such as DES and LSST. SDSS is an example that combined both approaches very successfully. The advantage of imaging surveys is that they can collect data from a very large number of objects relatively quickly. This is usually done using a relatively small number of filters (five or six) from which then a photometric redshift can be extracted. 
These redshift estimates have limited accuracy, which degrades the sensitivity of some cosmological probes.  More
  importantly, photometric redshifts suffer from occasional catastrophic failures -- misinterpretation of
  spectral features -- which lead to hard to quantify systematic uncertainties.
Spectroscopic surveys on the other hand are observationally very expensive and therefore the number of objects that can be observed, as well as the effective depth, is limited. In addition, spectroscopic surveys rely on good target catalogs. The very important advantage of spectroscopic surveys is that they deliver essentially perfect redshift measurements, providing high-fidelity 3D maps of the universe, which are often essential for large scale structure probes of cosmology. 

%
%

\subsection{Overarching Topics}

\subsubsection{Theory and Modeling} 

Theoretical investigations have helped initiate and shape the cosmic program. Ideas for the physical mechanism driving acceleration have come from theorists, as have many of the ways to extract information from the cosmos about these ideas. To mention just a few examples, using the BAO signature as a distance indicator, the power of weak lensing to probe dark energy, and the power of redshift space distortions to probe the growth of structure are all ideas that originated in theoretical papers. Inflation, the realization that gravitational waves would be produced during such an epoch, and the proposal to detect these via their unique signature in the CMB are all examples of theoretical contributions. A large challenge facing the theoretical community now is to understand the current epoch of acceleration both by developing consistent models of cosmic acceleration and by exploring optimal ways to connect these models to observations. The next few paragraphs touch on some of the most difficult issues involved in constructing viable models of cosmic acceleration, focusing on modified gravity models.



An example of a modification to gravity that can drive acceleration is {\it massive gravity}. The notion of giving the graviton a mass has been intriguing to theorists since Fierz and Pauli's discovery of a unique ghost-free linear Lagrangian for a massive graviton, but until recently progress had stalled due to a powerful no-go theorem 
describing the obstacles to finding a nonlinear completion of this theory. However, in the last few years, a loophole to these objections has been found, and a consistent interacting theory of a massive graviton 
has emerged. Interestingly, this 
{\it ghost-free massive gravity} admits self-accelerating solutions in which the de Sitter Hubble factor is of order the mass of the graviton. Since having a light graviton is technically natural, such a solution is of great interest in the late-time universe as a candidate explanation for cosmic acceleration. This is a good example of the difficulty of constructing a viable model, because even after the ghost-free model was discovered, theorists still needed to address a conflict with local tests of gravity. Massive gravity models do this by employing a screening mechanism, the {\it Vainshtein mechanism}, in which nontrivial derivative structures in the theory allow for varying field profiles around massive bodies, which lead to much weaker forces that one might have naively expected.

Many extensions of ghost-free massive gravity have been proposed, leading to a variety of cosmological solutions and other tests. It remains to be seen whether a completely satisfactory theory of massive gravity can drive the current epoch of acceleration, so these theories form an exciting frontier in theoretical research. Another reason for the excitement is that massive gravity is an example of a theory that allows us to view the cosmological constant problem in a different light. Rather than asking: {\it why is the observed cosmological constant so small?} we are able to ask the slightly different question: {\it why does the cosmological constant not gravitate very strongly?} and to take seriously the notion that the large cosmological constant generated via matter loops is physical, but that it does not strongly curve space-time. This phenomenon of {\it degravitation} is intimately tied to massive gravity, since any theory that exhibits this must reduce to a theory of massive/resonance gravity at the linearized level. Other models employ different screening mechanisms, invoke nontrivial derivative self-interactions, lead to sound speeds different than the speed of light or even effective equations of state less than $-1$. 


All of these approaches raise important questions of theoretical consistency, and high energy theorists, armed with decades of experience developing quantum field theory, are uniquely qualified to confront the challenge of constructing viable models of cosmic acceleration.

Parallel to constructing these viable fundamental physics models, an equally important part of the theory effort involves analytic and semi-analytic techniques to both connect these models to data, and to understand the extent to which there are relatively model-independent measurements that can constrain classes of ideas. Past examples of these approaches are 
parameterized post-Friedmann frameworks that capture departures from the predictions of the cosmological constant model;  
consistency tests between the redshift-distance relation as measured by SN or BAO and the rate of growth of structure that could distinguish between dark energy and modified gravity; 
probing differences between 
the dynamical potential and the gravitational lensing potential by comparing gravitational lensing to galaxy clustering; similar probes on intermediate scales by observing the dynamical infall of galaxies onto clusters; and on small scales, by comparing lensing masses to dynamical masses of clusters. 
We expect ideas like this to proliferate over the coming decade and to play a key role in answering the most important questions in cosmology.

A third component of theory is the ability to make predictions for observables even on small scales, in the nonlinear regime. 
The amount of information grows appreciably (e.g., the number of Fourier modes grows as $k_{\rm max}^3$) as one includes small scales in an analysis. 
Therefore, cosmological simulations, which include the relevant nonlinear physics, will greatly enhance the scientific output of upcoming surveys. 
Cosmological simulations then play the same role for surveys that Lattice QCD does for colliders: they enable physicists to extract the signal from the complex observations.

The accuracy requirements on simulations are daunting,
given the forecasts for the statistical errors expected from LSST and DESI. 
In order to be fully prepared, we need a concerted effort for modeling similar to the effort developed for high energy physics experiments.
For example, while many approximate methods have been
developed to make neutrino simulations computationally feasible, a
full understanding of the errors due to the approximations has not
been carefully developed on all scales. Similar challenges exists for models of dynamical dark energy, self-interacting dark matter, and those with features in the primordial power spectrum. 
A different (and very exciting) challenge will arrive if evidence emerges for physics beyond the cosmological constant. Currently, we
have no large, very detailed simulations for modified gravity models for several reasons: it is not clear
how to best explore different theories in a systematic way. Even if we solve this problem, our current approaches are computationally very
expensive. Again given the fact that most information is hidden on
small scales, developing a viable simulation approach is essential.

Simulations are also crucial for investigating astrophysical systematics in detail, such
as the effect of baryonic physics on the power spectrum. Currently, results from different simulations
vary considerably on small scales. Another major challenge is adding different types of galaxies at
different depths on top of the gravity-only large scale structure
backbone information. Again, a range of methods has been developed for this task
but so far, none of them is fully satisfactory in reproducing all of the
observational data. Finally, estimates of the
covariance matrix rely heavily on simulations. Currently, brute-force
methods for generating covariances from simulations are
computationally too costly, and new approaches for obtaining
covariances are being actively investigated. 



\subsubsection{Synthetic Sky Maps and End-to-end Simulations}

Beyond their contributions to theoretical efforts, simulation and modeling efforts play two other central roles in modern cosmological
surveys. Broadly speaking, simulations allow us to (i) generate synthetic sky catalogs at different wavebands for pipeline and algorithm validation, and (ii) construct end-to-end simulations of the full survey or parts of it. 

Synthetic sky catalogs serve several purposes. They allow for detailed validation
of analysis pipelines and algorithms. The important fact here is that
the {\it truth} is known and the catalogs can be used to carry out tests in 
controlled ways. They therefore also can form the basis for data challenges
to prepare the analysis teams for real data and understand if their analysis software is up to the task of extracting the subtle signals under investigation. In addition, they can be used to mimic realistic data rates from the surveys to ensure scaling of the analysis codes. 

End-to-end simulations of the full survey have started to become
possible; for example, LSST will attempt to have a full
modeling pipeline starting from simulated synthetic skies (as described above) to modeling
of the atmosphere and the telescope optics to the detectors in order
to generate images as close to the real observations as
possible. Other surveys have also carried out simulations of parts of
this pipeline.  These simulations are very important to understanding
systematics effects in the observations themselves that can be induced
due to the observational hardware, calibration issues, or data
analysis pipeline shortcomings. Detailed simulations allow us to
investigate those possible systematics and their effects on the data
and cosmological constraints and, eventually, to mitigate them.

\subsubsection{Photometric redshift calibration of LSST objects}\label{photoz}

LSST is critically dependent upon photometric redshifts --
estimates of the redshifts of objects based on flux information obtained
through broad filters -- for all probes of dark energy. High quality photometric redshifts can transform an inherently two dimensional map into a full 3D map of the universe, thereby harvesting much more of the information in the sky.

Higher-quality, lower-scatter photometric redshifts will result in smaller random errors on
cosmological parameters; while systematic errors in photometric redshift
estimates, if not constrained, may dominate all other uncertainties from these
experiments. The desired optimization and calibration is dependent upon
spectroscopic measurements for secure redshift information; a comparatively
modest investment in spectroscopic studies can greatly enhance the dark energy
constraints from LSST. 

Although a variety of methods are still under development and will be tested on the Dark Energy Survey, several robust conclusions are already possible. First, a {\it training set} of galaxies that trace those found in LSST is crucial; obtaining spectra for tens of thousands of faint objects found in LSST will require a large amount of telescope time but will improve the dark energy figure of merit by an order of magnitude. Current thinking is that these gains can be realized only by also {\it calibrating} the errors on the photometric redshifts. Rough first estimates are that DESI spectra will help considerably, but additional spectra in the South will improve the figure of merit by an additional 30\%.

\subsubsection{Cross-Correlations} 

The community is just beginning to explore the gains that can be obtained by combining data from different surveys. Here we list a few examples, but it cannot be emphasized enough that this is one of the most exciting topics in observational cosmology, one likely to progress significantly in the coming years.

\begin{itemize}
\item Gravitational lensing of background galaxies provides an estimate of the mass associated with foreground lenses, so calibrates the {\it bias}, the relation between the foreground galaxy density and mass density. Measurements of galaxy clustering then, at least on large scales, can be cleanly related to the clustering of matter, and it is the latter for which very accurate predictions are possible. Combinations of lensing and clustering therefore lead to tighter constraints on the growth of structure than either can independently. 
\item The CMB is also lensed, and over the past five years, CMB lensing has progressed from the first 4-sigma detection to an all-sky 30-sigma map by Planck. Future CMB projects will do much better, and the cross-correlation between a CMB lensing map and a galaxy sample in a given redshift slice is projected to provide yet more information about the bias of that sample, at the few percent level. Like lensing of galaxies, this information too will flow into reduced errors on the clustering of matter and therefore on cosmological parameters.  Overlap between CMB surveys and galaxy surveys therefore is extremely beneficial.
\item Regions over which we have both spectroscopy and shape measurements will be used to compare the lensing signal (one measure of the mass) with the velocity signal (another measure via the continuity equation). Dark energy models predict a unique relation between these two, while modified gravity models generically predict differences. This test has recently been carried out on SDSS data, but could progress to the percent level with future overlapping spectroscopic and imaging surveys.
\item The number of massive galaxy clusters could emerge as the most powerful cosmological probe if the masses of the clusters can be accurately measured. For this purpose, an arsenal of tools is building up: galaxy lensing, cluster velocity dispersion, CMB cluster lensing (with the first detection in 2015), galaxy richness, the distortions in the CMB due to scattering of hot electrons, and X-ray measurements. Cross-correlating and calibrating all of these with one another is a decades-long project, one that will require overlapping surveys at multiple wavelengths.
\end{itemize}

\subsubsection{Pixel Level Comparison}

While LSST is performing its decade-long deep, wide, fast survey of the sky, two major space missions will be operating contemporaneously:  Euclid, a European Space Agency medium-class mission, and WFIRST, a NASA-funded medium space mission ranked as highest priority in the last decadal survey.  LSST, Euclid, and WFIRST have all been designed to use similar techniques to constrain the properties of dark energy and gravity, but their detailed properties are rather different, and in many ways, they are largely complementary.  This implies that an organized combined analysis of data from all three projects has the potential to greatly increase the science return over what can be obtained through independent analyses of the data from the separate projects themselves.  

Euclid utilizes a 1.2 m anastigmatic telescope with two primary scientific instruments:  A visible imaging array (VIS) with a single broad-band filter, and a near-infrared spectrometer and photometer (NISP), which will perform imaging in 3 bands (Y, J, H) and grism spectroscopy in the 1-2 micron wavelength range.  The VIS instrument will provide high spatial resolution (0.2 arc sec) imaging for weak lensing investigations, while NISP will produce high quality infrared galaxy photometry and low resolution spectroscopy for both weak lensing and large-scale structure studies.  Euclid is designed to survey 15,000 square degrees, comparable to the LSST survey area, but with considerably lower sensitivity.

WFIRST will utilize an existing 2.4 m telescope coupled to a wide-field infrared imaging array and an integral field unit spectrometer, as well as a coronagraph instrument.  The wide-field imaging and spectroscopic capability operates with multiple bands between 0.7 and 2 microns in wavelength.  The pixel scale for the imager is 0.11 arcsec.  WFIRST will achieve comparable depth to the full LSST ten-year survey, but over a significantly smaller fraction of sky, $\sim$2400 square degrees.

As indicated above, LSST, Euclid, and WFIRST are highly complementary.  LSST and Euclid will cover large fractions of the sky, but at different depth, with different spatial resolution, and in different wavelength bands.  WFIRST will get to comparable depth to the full LSST with much higher spatial resolution, but in the near infra-red and over a more limited region of the sky.  The biggest advantage of combining data from these three missions will come from significant reduction in the systematic errors that affect each of the different probes differently.

For weak lensing and large-scale structure measurements, LSST is essential to provide the visible band photometry necessary to estimate photometric redshifts, when combined with the infrared colors from both Euclid and WFIRST.  For LSST photometric redshift determination, the additional infrared data will help to narrow the error distribution and suppress catastrophic errors.  In addition, shear comparisons over limited regions of sky between the higher spatial resolution images obtained by Euclid and WFIRST can help to calibrate both multiplicative and additive errors in the LSST shear determinations.  Finally, the higher resolution imaging provided by the space missions will aid in galaxy deblending in the LSST image analyses.

A combined analysis of LSST, Euclid, and WFIRST data will also be very important for strong lensing investigations.  The higher spatial resolution images will allow accurate determination of the lens model, while LSST will provide the time-domain information necessary for the determination of time delays between multiple images.  For Type Ia supernova investigations, the broader spectral band afforded by the infrared imaging extends the redshift range that can be studied considerably.  With its larger survey area and large number of repeat visits, LSST is optimally suited for SN discovery, however WFIRST will be better for obtaining precision spectrophotometric time series and for SN redshift determination.

While some of these coordinated analyses can be performed using catalog data from the three missions, much of the cosmological work will require a joint pixel-level common data reduction and, for supernova work, near-real-time coordination.  This is particularly true of galaxy photometry, especially given the complexities of galaxy deblending.  At present, such work is beyond the scope of what is currently budgeted for the three individual missions, a consequence of the fact that Euclid is European-led, while LSST and WFIRST are American-led, and the fact that LSST construction is supported by NSF and DOE, while WFIRST construction is supported by NASA.  Therefore, it is important that a cross-funding joint program or several bilateral programs are established to fully ensure we are able to get the optimal science return from the ``system" represented by these three very powerful facilities.

\subsection{Supernovae}

Type Ia supernovae have a long track record as a cosmological tool: they provided early evidence for cosmic acceleration (resulting in a Nobel Prize!), and today are included in all of the tightest constraints on the dark energy equation of state. Going forward, there are numerous opportunities to improve on supernova cosmology, ranging from currently planned facilities (e.g. LSST and WFIRST), to new programs on optical multi-fiber spectrographs, to new technologies (sky line suppression).

\subsubsection{Follow-up spectroscopy of LSST supernovae and hosts}

LSST is projected to discover millions of supernovae, including 200,000 well-sampled Type Ia SNe in the wide field survey and ten of thousands of Type Ia SNe with densely sampled light curves in the deep drilling fields. For reference, the current state-of-the-art instrument is the Dark Energy Survey, which is projected to find $\sim 3000$ SNe. To place the LSST-discovered supernovae onto a Hubble diagram and extract cosmological constraints, follow-up spectroscopy will be essential at high-redshift, and further anchor observations will be required at low redshift. The potential gains are enormous: even in 2030, a reliable Hubble diagram drawn from 30,000 Type Ia SNe will be a signature achievement.

Obtaining spectra of all densely sampled Type Ia SNe detected in the deep drilling fields is likely to be too expensive, so we envision a sample of thousands of spectroscopic Type Ia SNe. One could obtain follow-up spectroscopy of the host galaxies of the remaining SNe later with a multi-fiber spectrograph to provide the host galaxy redshift and type, and help break degeneracies in supernova classification. This photometric sample would have larger scatter than the spectroscopic SNe detected by LSST, and would be at lower redshift than the WFIRST SNe, but will be much larger.

Spectrophotometric observations covering optical wavelengths from the Nearby Supernova Factory have proven effective at producing a small scatter on the Hubble diagram. The 2--3$\times$ smaller scatter achieved using spectrophotometric ``twin'' SNe already sets an impressive limit on the remaining systematic uncertainty due to dust extinction and astrophysical diversity among SNe Ia. These data indicate that many of the problems ascribed to dust extinction as measured from broadband imaging arise in part from un-modeled diversity in strong SN spectroscopic features that directly distort the color, and hence the inferred dust reddening, from broadband imaging. Such issues could be addressed at moderate redshift by spectrophotometric observations of a subset of SNe discovered by LSST.

\subsubsection{Near infrared studies}

Another key to advancing supernova dark energy studies is to include measurements in the near-infrared region (NIR, from 0.9 to 2.5 microns), which would both enhance the sample of SN at higher redshifts and also probe the impact of dust in the host galaxy. The large SN sample expected from LSST will require corresponding improvements in the understanding of small systematic effects. Dust in the host galaxy is particularly daunting, and since longer wavelength light is less sensitive to small dust grains, NIR measurements are one avenue to reduce this systematic. In addition to reducing dust systematics and discovering high redshift SN, rest-frame NIR observations provide smaller raw scatter than optical measurements. At $z>1$, the rest-frame optical becomes the observer-frame NIR, and so NIR observations are key to this redshift range. Historically, NIR observations have been complicated by the extremely bright (and variable) airglow emission, but at least two developments could dramatically increase SN follow-up capabilities in this spectral range.

The WFIRST mission, which will launch in 2024 and hence operate during LSST's 10-year observing program, will carry a wide-field camera and integral field spectrograph sensitive out to $\lambda = 2$ $\mu$m; by being above the atmosphere, it will avoid the airglow problem entirely. As originally conceived, WFIRST would both detect and follow up its own supernovae in a self-contained program, however there would be several benefits to a coordinated program between LSST and WFIRST. LSST could provide WFIRST with observer-frame optical imaging, including SN discoveries out to $z \sim 0.8$; in turn, WFIRST could provide spectrophotometry sufficient to apply the twin SN technique and study dust, and observer-frame NIR imaging for additional dust studies, for a large subset (2000 or more) of active LSST SNe.

It is also possible that new technology will allow suppression of NIR airglow from ground-based observatories, since most of the emission is in a forest of discrete lines but the SN spectrum is continuous. There have been previous attempts at technology solutions for absorbing the specific airglow wavelengths while leaving the continuum intact, but these have not been scalable to the level required to meet the needs for SN cosmology. However, as detailed in the accompanying Technology document, there is an R\&D program on a ring resonator device for airglow suppression. If successful, this would open an opportunity for a ground-based observer-frame NIR instrument (combining adaptive optics and airglow suppression) that would greatly enhance the SN capability of LSST.

\subsubsection{Low-$z$ supernovae}

Finally, for moderate and high-$z$ SN measurements to be used most effectively for the cosmological measurements there is still a complementary need for further low-redshift anchor SNe to be measured at high $S/N$ using well-calibrated instrumentation.  Ideally, this data set should be spectrophotometric in the optical (similar to the Nearby Supernova Factory dataset), but additionally have near-IR observations (photometric NIR, if not spectrophotometric).  The additional data set would contain at least 800 more SNe in the nearby Hubble flow. This could be obtained with existing nearby SN searches, or it could be obtained by following LSST low-redshift SN discoveries if the LSST search were set up for that purpose. In either case follow-up instrumentation (IFU and NIR) and telescope time would be needed.


\subsection{Weak Lensing} 

Weak gravitational lensing is a powerful cosmological tool, since the observables are directly related to the matter perturbations (or, in the case of modified gravity, the gravitational potential) of the universe. It is also technically challenging: the small lensing-induced shear signal is easily masked by systematic effects; the intrinsic spread in galaxy shapes requires large samples; and spectroscopic redshifts are typically available for only a tiny fraction of the sources. In response to this challenge, the community has devised an extensive Stage IV weak lensing program. This includes LSST, which will obtain deep 6-band optical imaging over most of the Southern sky over the course of its 10-year observing program; WFIRST, which will provide deep 4-band infrared imaging at much higher resolution, but over a smaller area; and Euclid, which will provide a single optical band and shallower infrared photometry, with greater area coverage but less internal redundancy than WFIRST. Together, these surveys will be extremely powerful, but will still require calibration of their photometric redshifts (see \S\ref{photoz}). In addition, dedicated work on algorithm development for extracting galaxy shapes is being carried out within DES and LSST. This kind of work is essential not only in weak lensing but for many of the probes in order to maximize scientific output from investments in the projects.

In the future, a range of techniques may enable further gains using weak lensing. One exciting possibility is the kinematic weak lensing method.

\subsubsection{Kinematic weak lensing}

Measurement of the lensing shear acting on a galaxy is usually dominated by two types of noise -- {\em measurement noise}, which is the result of photon counting statistics with a finite aperture and exposure time, plus noise associated with the detectors; and {\em shape noise}, which describes the intrinsic deviation of a source galaxy from roundness. As exposure times increase, the measurement noise decreases, but the shape noise is usually treated as irreducible and is usually a factor of ten to a hundred larger than the signal. Kinematic weak lensing proposes to reduce shape noise significantly by using measurements of the velocity structure of disk galaxies to learn about the intrinsic shape of the galaxy. For a perfect disk, the radial component of the velocity is greatest at the position of the true (i.e. pre-shear) major axis, and is equal to the systemic velocity at the position of the true minor axis. Deviations from the ideal case occur, and represent a ``shape noise'' for kinematic weak lensing, but at a lower level than the raw shape noise.

The principal data set required for kinematic weak lensing is a 3D map ($x$ position -- $y$ position -- wavelength) of the emission lines for each galaxy, which could be obtained with an integral field spectrograph (e.g. with a deployable fiber bundle) or, at some cost in observing time, by stepping a fiber or slit across the target galaxy. The spectral resolution must be adequate to resolve typical ($\sim 200$ km/s) internal velocities in the target galaxies, and -- just as with regular weak lensing -- the galaxy must be spatially resolved. 

Kinematic weak lensing carries a number of advantages relative to traditional weak lensing: in addition to suppressing shape noise (and thus beating a ``fundamental'' limit for traditional methods), the measurement also returns a spectroscopic redshift for each source galaxy (eliminating the photo-$z$ problem) and solves for the intrinsic orientation of the galaxy, thus potentially reducing another lensing systematic problem: that nearby galaxies are intrinsically aligned. Its principal disadvantage is that the investment of telescope time to obtain high-quality spectra is much greater than the time required for multi-band imaging. Studies of the science reach of kinematic weak lensing as a function of telescope aperture, multiplexing capability, and other parameters (e.g. fiber size, target selection) are underway. Another area that needs work is understanding how the demanding systematic requirements of traditional weak lensing measurements (e.g. point spread function determination) map onto the kinematic weak lensing approach.


\subsection{Clusters}

The primary challenge in Stage IV cluster cosmology will be to estimate cluster masses utilizing the combined strengths of optical, X-ray and
mm-wave observations.
The most promising route for absolute mass calibration is weak
gravitational lensing. At low redshifts ($z<0.5$), optical measurements
of weak lensing of background galaxies by clusters has already been
demonstrated to provide accurate mass calibration with well controlled
systematics at the 5-10\% level. The challenge will be to extend this work to cover the full
redshift range spanned by the cluster population ($0<z<2$) at the
precision necessary 
to reliably extract information about physics.
Improving photometric redshifts by obtaining tens of thousands of spectra as outlined in \S\ref{photoz} will significantly improve weak lensing mass estimates and therefore is perhaps the most important step towards this goal.

In addition, precise relative cluster mass estimates for individual systems are provided by targeted X-ray
measurements of the gas mass and gas temperature of clusters. These
X-ray measurements correlate more tightly with halo mass than the
survey observables used to identify clusters, and thus can provide
improved information on both the shape of the mass function and the
intrinsic scatter between halo mass and the survey observables. 

Another promising application is to take spectra of many galaxies in or behind a cluster. 
This would enable phase space modeling of the clusters, combining dynamical data with lensing data.  A spectroscopic follow-up program along these lines has already been carried out on clusters identified in CMB experiments, with of order a few thousand spectra of galaxies in of order a hundred clusters. The spectral requirements are not prohibitive since there is intrinsic scatter due to projection effects, so data from many clusters will have to be averaged together. But, clearly, with LSST poised to discover tens of thousands of massive clusters, many of which will be at redshifts too high to obtain accurate weak lensing masses, a program that measures the velocities of, e.g., a million galaxies in clusters would be
extremely interesting. The accurate cluster spectra will also be useful for mapping the cosmic velocity field, especially when combined with CMB maps, which are sensitive to the velocity via the kinetic Sunyaev-Zel'dovich effect. Finally, spectra of background galaxies will help not only with weak lensing by shapes but also with estimates of cluster mass via magnification.

\subsection{Baryon Acousic Oscillations and Redshift Space Distortions}

\subsubsection{3D Clustering Beyond DESI}

Spectroscopic surveys like DESI will provide a very detailed measurement of BAO out to $z=1.5$. Nonetheless, for reasons outlined in \S\ref{information}, even after DESI completes, there will remain many ways to gain information from measurements of galaxy clustering. As long as the number density of detected objects is sufficiently large on the BAO scale (larger than $1/P$, the inverse of the power spectrum), 
a survey will extract most of the BAO information. DESI is designed to achieve this at $z<1$ and it is thus unlikely that there 
can be much improvement with a survey that measures more faint objects (a {\it deeper} survey) at these low redshifts: improvements are possible only by adding more volume. 
One possible strategy is to go to higher redshift (with potentially 
significant increase in volume); another is to survey a different area in the sky (where the volume increase is more modest, typically a factor of 2).

For redshift space distortions (RSD), the power spectrum on the relevant scales is smaller so the required density is larger. Therefore, besides the large amount of information available at high redshift, 
improvements with a deeper survey at relatively low redshift could be significant and both options should be considered. 
For a deeper survey with a higher number density the corresponding bias is lower, but due to the increase in the number density the overall signal to noise is 
typically higher. Moreover, the lower bias makes the angular anisotropy induced by redshift space distortions 
larger. This angular anisotropy generated by the velocity field, which traces the perturbations in the overall matter density, is what  
allows one to determine the power spectrum. Because of the larger angular modulation
of lower bias tracers, the corresponding errors on the matter power spectrum are smaller. 
For example, DESI observations over the same redshift range as BOSS are expected to yield a factor of 2 smaller errors due to the 
fact that DESI observes emission line galaxies, with a lower bias than luminous red galaxies observed by BOSS. 

A second argument for observing lower mass and bias galaxies is that one is tracing the dark matter better. Galaxies are 
formed at the centers of dark matter halos, which form at the peaks of a smoothed initial dark matter density field. The smoothing 
increases with the halo mass. Clusters are expected to erase all information for $k>0.5h/Mpc$, while 
the corresponding scale for regular galaxies exceeds $1Mpc/h$. Higher number density of galaxies also traces the dark matter 
better. If one knows the host halo mass one can weight the galaxies such that the resulting field can reproduce 
the underlying dark matter, with higher number density samples reproducing it better. 
As a result one expects that higher density tracers can lead to a better reconstruction of the dark matter initial conditions. 

A third argument for increasing the number density is the multi-tracer approach. On large scales one is usually limited by the
sampling variance, which is the finite number of the gaussian random modes one is measuring. However, by taking a ratio between 
the tracers of a different bias, such that the randomness of the mode cancels out,
one can extract certain types of information without the sampling variance limitation. Two examples of this are geometry 
information 
and primordial non-gaussianity. In both cases one expects significant improvements with this 
technique. 

In short, 3D clustering beyond DESI promises significant improvements in the amount of information about our universe.


\subsubsection{Probes of the Intergalactic Medium} 

The Inter-Galactic medium (IGM) usually refers to the baryons, mostly primordial hydrogen and helium, in the space between galaxies. Like galaxies, the IGM traces the underlying dark-matter fluctuations but is in many respects easier to understand theoretically. Galaxies form in very over-dense regions and their formation is governed by highly non-linear hydrodynamical, chemical and nuclear processes of star formation and evolution. The IGM, on the other hand, is comprised of modestly over- and under-dense regions. On large scales, baryons follow dark matter and on small scales baryons are smoothly distributed, supported by gas pressure. The IGM therefore offers significant advantages over galaxies as a tracer of structure: it can often be modeled a-priori and is quasi-linear to significantly smaller scales. It can be used to constrain free-streaming length of the dark matter in ways that galaxies cannot.
However, since the IGM is not luminous, it is difficult to measure directly. Fortunately there exist several promising techniques.

\newcommand{\lya}{Lyman-$\alpha$}
In the \lya\ forest technique, distant quasars are used as back-lights illuminating the IGM. Light is scattered by the neutral hydrogen in the IGM though the \lya\, line, resulting in density fluctuations of the neutral hydrogen being imprinted in the spectra of distant quasars. With the advent of BOSS, the number of quasars measured on the sky became sufficient to allow \lya\ forest to become a true three-dimensional tracer of the underlying structure by measuring correlations across single lines of sight.  This allowed measurement of the BAO feature from the \lya\ forest for the first time. In addition, BOSS measured BAO from the cross-correlation between quasars and the forest.  BOSS measured around 160,000 high redshift quasars and DESI will have over 600,000 high redshift quasars. These numbers could be improved by a factor of around two, but larger gains will require different sources.  One possibility is to use galaxies as back-lights, although they are likely to be noisier due to the larger intrinsic variability of galaxy spectra. 

Large-scale clustering of the \lya\, forest has not yet been measured with sufficient systematic control to measure shape of the three-dimensional matter power spectrum, but analyses are currently under way with both theoretical and experimental advances. Further research, code and simulation development is required before this becomes reality.

The IGM not only absorbs and rescatters the \lya\, light, but also faintly glows in the rest-frame \lya\, emission. This emission was first detected using BOSS data. The signal is very faint and contaminated by emission from galaxies below the survey's detection threshold and by interloper line emission. Therefore, it is likely that this signal can be used only in cross-correlation. 
However, for future integral field spectroscopic surveys, there could be significant
gains, since all the fibers, including majority of those looking at
spectroscopic targets, can be used for this technique.


\subsubsection{Neutral Hydrogen Surveys: 21 cm line}

The hyperfine splitting of neutral hydrogen is $\sim 10^{-6}$ eV, with a wavelength of 21 cm. Detecting this radiation at $\lambda=21\,(1+z)$ cm opens a new window on the distribution of structure in the universe,
measuring the neutral hydrogen density as illuminated by the CMB. The signal can be either in absorption or
in emission depending on whether the CMB temperature is larger or smaller than the {\it spin temperature} determined by the ratio of the density of the two levels.
Interestingly, the signal is expected to change sign at least once in the early history of the 
universe. One can observe both the mean 21 cm emission signal at a given redshift or its fluctuations. There are planned 
experiments to measure the former at higher redshifts ($z>6$), where the signal is strong. The latter however 
contains most of the information of cosmological interest and will be the focus here. 

There are several epochs when the signal may be observable and of interest to cosmology. 
The earliest one is before any stars have formed, $z>25$. The signal in this era is simple to model and
is proportional to the total gas density fluctuations.
On the experimental side, the signal during this epoch is weak when compared to 
the very large foreground contamination signal. 
As a result there are currently no planned observations in this regime, but the promise for cosmological studies
is large due to the large number of modes in the linear regime. 

The next epoch is before reionization, $z>15$, but 
after first stars ($z<25$). An overdense region initially {\it absorbs} CMB radiation because the spin temperature is tightly coupled to the gas temperature, which is relatively low. At some point though, the gas is heated by early sources of X-rays and the signal changes to emission. 
While for homogeneous heating or cooling the signal still traces the underlying gas density, 
there are astrophysical effects that complicate this simple picture during this 
epoch. The extent to which cosmological information, especially that from gravitational lensing, can be extracted in the presence of these effects is 
worthy of further study.

The signal during reionization, $6<z<15$, is characterized by large reionization bubbles inside which gas is 
ionized and is thus mostly without the neutral 21 cm signal, but outside of which there is 21 cm emission. These bubbles 
start small, 
grow in time and finally overlap, at which point the signal contrast disappears. 
Three dimensional maps of the neutral hydrogen during this epoch contain much information about cosmological parameters, but current thinking is that most of this will be obscured by astrophysical uncertainties. Path finder experiments over the coming decade could change this view. 
There is however one measurement during this epoch that is of 
interest to cosmology, which is the measurement of the reionization optical depth $\tau$. This can be obtained
by connecting the bubble fluctuation signal to the mean 
ionization fraction, and integrating it as a function of redshift. In the context of simplified parametrized models
it can constrain the reionization optical depth $\tau$ 
more precisely than 
CMB polarization. 
Since $\tau$ is degenerate with cosmological parameters such as the amplitude of fluctuations and 
neutrino mass, such a measurement would enable one to reduce the expected error on these cosmological parameters. 
For example, an experiment like HERA (Hydrogen Epoch of Reionization Array) combined with CMB-S4 polarization lensing
may potentially reduce the error on the sum of neutrino mass from 19 meV to 12 meV, and a 
future 21 cm mission could further improve upon this. 

A promising epoch for 21 cm cosmology is at lower redshifts, after the end of reionization ($z<6$). 
One reason is that foreground contamination is smaller at low redshifts, and at $z=1$
one only needs to reject 60\% of the data even if one does not attempt any foreground cleaning. 
The signal during this epoch comes from small pockets of high density 
neutral hydrogen inside dark matter halos. In many ways this signal is similar to 
galaxies tracing large scale structure: the signal is that of a biased and essentially discrete tracer of dark matter, with 
the amplitude of fluctuations on large scales determined by the product of the 
bias and the mean density of neutral hydrogen. 
As with galaxies, the power spectrum contains valuable information in the redshift space distortions, but extracting information about the growth of structure from these will be more difficult
since the mean density is not known. Experiments over the coming decade will test ideas, such as using the higher-point functions and cross-correlations, for extracting this information.

A second, and completely independent, method for extracting cosmological information, is through weak 
lensing  of 21 cm structures by the matter distribution along the line of sight. 
On large scales the effect is analogous to the effect of lensing on cosmic microwave 
background, and lensing generates a 4 point function of 21 cm intensity from which one can reconstruct the 
underlying projected matter density. 
On small scales the effect is analogous to the magnification bias effect on galaxies, except that one is 
adding all galaxies together, weighted by their HI luminosity. 
A major advantage over the magnification bias studies in galaxies is that the redshift information is 
essentially perfect. 
Due to the good redshift resolution of 21 cm signal one can attempt a
tomographic decomposition of the lensing signal into redshift bins. Moreover, the signal can in principle 
be obtained regardless of the underlying 21 cm emission model, although the signal predictions vary significantly 
depending on the model. The information content grows with the number of 21 cm modes
measured with signal to noise above unity and the predictions for SKA from the $z\sim 2$ signal suggest 
high levels of detection 
may be possible to achieve, although no fully realistic SKA simulations exist 
at present. Experiments over the next five years will test this idea.

Bias and shot noise of neutral hydrogen at $z>2.5$ remain major uncertainties: while high bias increases the signal, 
it is possible that the sources are so rare that the shot noise is a major issue for signal to noise.  
Foregrounds are a major source of signal contamination. 
For example, without any foreground cleaning one is forced to reject a wedge of Fourier modes 
with 97\% of the data around $ z \sim 8$, with the number increasing with redshift. 
Foreground cleaning of point sources into the wedge requires longer baselines and 
precise beam sidelobe measurement, and this remains a major source of uncertainty in current predictions. 


\subsection{Novel Probes}\label{novel}

 \begin{wrapfigure}{r}{0.35\textwidth}
  \begin{center}
    \includegraphics[width=0.3\textwidth]{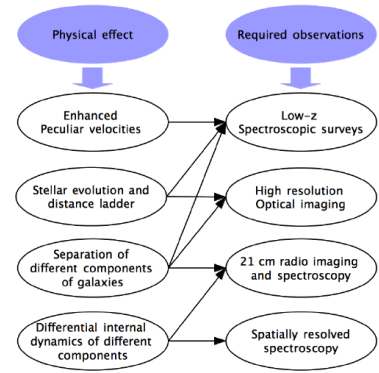}
  \end{center}
  \caption{Effects induced by difference of the gravitational force in screened and unscreened regions (left), and the observations (right) sensitive to these differences.}
\label{fig:novel}
\end{wrapfigure}
A suite of novel probes can discover signatures of modified gravity. As mentioned above, theories that go beyond general relativity in order to explain the current epoch of acceleration inevitably introduce new degrees of freedom, and these leave their imprint on a variety of scales. 
The most robust signature of modified gravity models is screening; the notion that since tests have established the validity of Einstein?s theory in the laboratory and in the Solar System, there must be a mechanism that renders the additional degrees of freedom ineffective in such regions. Novel probes search for the differences between the behavior of gravity in screened and unscreened regions.

These tests rely on stars, galaxies and black holes, and operate on much smaller scales than the cosmological tests mentioned so far, but can be even more powerful in testing screening mechanisms. 
Applications of astrophysical gravity tests have relied on archival data to date. The next big advance could come from mini-surveys designed for such tests, which can be carried out over a 5 year timeframe (the SDSS IV MANGA survey is currently being used for such an analysis). The mapping of the observations to gravity tests is depicted in Fig.~\rf{novel} taken from the Snowmass {\it Novel Probes} document. 
The observations needed are very diverse; here we mention two that require spectroscopic facilities and appear promising. 

LSST will detect a large number of faint galaxies that are relatively close to us (e.g., within 100 Mpc). Spectra of these galaxies would enable a simple test of how rapidly individual galaxies are infalling into potential wells associated with nearby galaxy groups or clusters. These infall velocities should differ at the ten percent level depending on whether the regions are screened or unscreened. A multi-fiber spectrograph, even on a 2-4 meter telescope, could estimate the infall velocities of hundreds of galaxies within 10 Mpc of a potential well. Repeating this in $\sim100$ regions would constitute a powerful test of gravity. 
This same program of probing the relation between potential wells and infalling matter can be carried out on groups and galaxy clusters at cosmological distances. LSST will enable the determination of the masses of clusters via gravitational lensing, so 
a spectroscopic follow-up program that probes the infall velocity of galaxies would also be sensitive to screening mechanisms.
This program will also require multi-fiber spectrographs but probably mounted on larger telescopes.

\section{Projects}\label{sec:project}

Many of the activities described above could be supported as part of the DOE research program, and they will likely lead to important enhancements to the science of DESI and LSST. Support for theory, simulations, and ring resonators for example could flow from modest R\&D investments and judicious use of DOE computing facilities. Multi-agency and multi-national agreements could help promote pixel level comparisons, and siting of CMB-S4 to overlap with large optical surveys would enable many of the gains mentioned in the cross correlation section.

Others mentioned above would benefit greatly from a new project. Below we list a range of possibilities, each of which extends the current program in significant ways and would enable tighter constraints on fundamental physics parameters. None of these is yet ready for CD-0, but 
at least one of these could go forward on a relatively short time scale at relatively low cost. The exception is the final project: its time scale and cost are larger than the others.

\begin{framed}
{\bf Ideas outlined in the previous sections lead naturally to a range of projects, at least one of which could begin construction in the 2020's.}
\end{framed}

\subsection{Optical Fiber Spectrograph on Existing Telescope, evolutionary from Existing Technology}

Many of the gains outlined in the previous section could be obtained with instruments on a telescope as small as 4m, but DESI alone in its first 5 years will be devoted to its main science goals so will not serve this purpose. The idea that there is much more to do than can be done with 5 years of DESI leads to the possibility of another spectroscopic survey, dedicated to the wide range of science outlined above. Two possibilities are outlined in \S5.1.1 and \S5.1.2.
It is important to emphasize that these two possibilities may have an advantage beyond a simple factor of 2 even if the instrumentation and time allocation are similar to DESI. The intriguing idea is that a future spectroscopic survey could target objects with good photometric redshifts. The much narrower range of possibilities for the redshift translates into a much smaller probability of obtaining an incorrect redshift. Therefore, lower signal to noise is required and, most importantly, the time to obtain a single redshift could be reduced significantly. Early estimates in reduction in time required per object range from 3-5. Further research into this idea will help solidify the scientific scope and case for each of these two projects.

\subsubsection{New Instrument in the South}

A Southern spectroscopic survey facility would improve the scientific capabilities of all the probes of dark energy. Spectra of even a small sample of LSST galaxies would reduce the uncertainties associated with photometric redshifts, thereby improving all the probes. 
Such a survey could obtain spectroscopic redshifts of tens of thousands of active Type Ia SN, making it by far the largest sample of this type even in $\sim$2030.
If the kinematic weak lensing idea works out, a Southern spectroscopic survey could be the perfect shear calibrator for LSST and, again depending on how well the technique works, even serve as a pristine, shape-noise-free lensing survey. Cluster science would benefit not just from the improvement in photometric redshifts but also from accurate cluster velocities and velocity dispersions. With flexibility, such a survey could enable an world-leading program of Novel Probes. Finally, using LSST targets with photometric redshifts, as described above, such a survey could obtain spectra for many more galaxies than will DESI, thereby promising strong improvements in BAO and specially in RSD.

\subsubsection{DESI-2}

Running DESI in the North past its current expected end date of 2024 is a natural possibility with significant scientific benefit and low cost.
Like the first possibility, the exact nature of a DESI-2 survey will be defined by developments over the coming 5 years. Despite the Northern location, there is considerable overlap on the sky between DESI and LSST. So, for example, while DESI North will not be able to take spectra of all LSST faint objects, it could obtain redshifts for more targets in its second 5 years than in its first 5 years by using LSST targets in the region of overlap. While it could not capture all LSST supernovae, it could likely greatly enhance the sample. Indeed, the scientific case for spectroscopy is so overwhelming that running spectroscopic surveys in both the North and South seems very compelling.

\subsection{Low Resolution Spectroscopy of a Billion Objects}

A multi-object spectrograph is a natural way to extend DESI: obtaining more objects for 3D maps. There is another possibility for a project to extend the Stage IV program: obtaining low resolution spectra for a substantial fraction of LSST objects. A number of possible technologies are under consideration, including MKIDs, narrow band filters, or even photometric redshift improvement by shifting the LSST filters. The scientific gain from a project like this has yet to be fully quantified, but efforts are underway to do this, and similarly there is a strong R\&D effort in progress to improve the spectral resolution of MKIDs. For example, current estimates assuming continued successful development of MKIDs are that spectra could be obtained for a billion LSST-detected galaxies in ~3 years on a 4m telescope.

\subsection{21 cm Survey}

A new dedicated 21 cm instrument with an optimized antenna array,
larger collecting area, and sufficient frequency coverage could significantly improve on measurements of large scale structure, extending them to 
higher redshifts (either before or after the epoch of reionization). It could also probe weak lensing distortions of fluctuations 
by structures along the line of sight.  None of the currently planned experiments are looking into this regime, but the 
likely configuration could be similar to the Canadian Hydrogen Intensity Mapping Experiment (CHIME) with more collecting area -- of order $(200m)^2$ -- to obtain higher redshift (lower frequency) information.
Many of the technical issues are in the sweet spot of DOE capabilities and are outlined in the accompanying Technology document.



\subsection{High Resolution Spectroscopy of a Billion Objects}

A most ambitious project would be one that obtained high resolution spectra of a large fraction of LSST objects. Such a {\it Billion Object Apparatus} (BOA) would come close to attaining the parameter improvements depicted in the right panel of Fig.~\rf{dpower} and open up many avenues for new discoveries. Here we outline some of the requirements, and these form a natural segue to the accompanying Technology document.

The DESI spectrogram will take spectra of 5000 objects using 10 spectrographs with 500 traces each. The Prime Focus Spectrograph (PFS) instrument is similar and the Maua Kea Spectroscopic explorer is somewhat more ambitious. Assuming a modest improvement in technology to a 2000 trace-spectrograph taking tens of thousands each, it seems that tens of thousands of traces simultaneously should be technologically feasible. An important improvement would be to extend the redshift range into the red using new Ge CCDs. The spectral resolution requirement should be studied more carefully but likely the current sweet-spot of around $R\sim2000-4000$ should continue to be sufficient.

It is unlikely that the DESI model with robotic fiber positioners will scale to this task. Making mechanical positioners smaller is difficult: a solid state solution would be preferable. There are several options available, all of which would require further R\&D. One possibility is to route light around using an image slicer and/or micromirror arrays. 
Perhaps the most compelling strategy would be to project thousands of fibers along the same spectroscopic trace and only let the light pass through selected fibers using microshutters or an equivalent strategy. This would allow one to fill the focal plane with fibers at packing efficiency around 50\% and then use dithering algorithms to account for dead space.

As the number density of fibers increases, there will inevitably be fiber positioning inefficiencies in any system with a finite spectral trace ``patrol radius'' (including current DESI and PSF designs). However, large number of fibers that are not taking object spectra could be used for Lyman-$\alpha$ (and other line) intensity mapping of the IGM. Moreover, even for spectra with objects on them, since the noise is dominated by the sky noise, their statistical utility after the model for the targeted object is subtracted should be roughly the same (but there might be subtle systematic effects associated with this procedure). 

\newpage
\begin{center}
{\bf \large Appendix}
\end{center}
\appendix
\section*{Process and Input}

The Cosmic Visions Dark Energy group was formed in August 2015. Between that time and the end of January 2016, the group of eight members held weekly telecons. There were also monthly telecons with Kathy Turner and Eric Linder from
the Office of High Energy Physics, DOE Office of Science
that were attended by management of the five HEP labs. Representatives of the group met with leaders of DESI at their November collaboration meeting and with leaders of the LSST Dark Energy
Science Collaboration at their October collaboration meeting. There were three workshops held to gather input for the white papers:
\begin{itemize}
\item Brookhaven, October 1, 2015. Agenda and slides available at \href{https://indico.bnl.gov/categoryDisplay.py?categId=124}{BNL}
\item Fermilab, November 10, 2015. Agenda and slides available at \href{https://indico.fnal.gov/conferenceOtherViews.py?view=standard\&confId=10639}{FNAL}
\item SLAC, November 13, 2015. Agenda and slides avaiable at \href{https://indico.fnal.gov/conferenceDisplay.py?confId=10842}{SLAC}
\end{itemize}
The group of eight met for a one day face to face meeting on January 14, 2016 at Fermilab.

These white papers freely used written input from the following contributors: Chris Bebek, Daniel Eisenstein, Juan Estrada, Brenna Flaugher, Eric Huff, Bhuvnesh Jain, Steve Kahn, Steve Kuhlmann, Chris Leitz, Michael Levi, Adam Mantz, Ben Mazin, Jeff Newman, Paul O'Connor, Aaron Parsons, Jason Rhodes, Eduardo Rozo, Will Saunders, David Schlegel, Risa Wechsler, and Michael Wood-Vasey.
In addition, we held numerous conversations with community members including: Brad Benson, Lindsey Bleem, Brenna Flaugher, Kyle Dawson, Tom Diehl, Josh Frieman, Salman Habib, Bhuvnesh Jain, Eduardo Rozo, Chris Stubbs, Martin White, and Michael Wood-Vasey.



\end{document}